# Anisotropy of the magnetic properties of the $FeTe_{0.65}Se_{0.35}$ superconductor

S. I. Bondarenko and O. M. Bludov

*B. Verkin Physics and Technology Institute for Low Temperatures, NAS Ukraine, 47 Lenin Ave., Kharkov 61103, Ukraine*

A. Wisniewski and D. Gawryluk

*Institute of Physics, Polish Academy of Sciences, Aleja Lotnikow 32/46, Warsaw PL-02-668, Poland*

I. S. Dudar, V. P. Koverya, V. Yu. Monarkha, A. G. Sivakov, and V. P. Timofeev

*B. Verkin Physics and Technology Institute for Low Temperatures, NAS Ukraine, 47 Lenin Ave., Kharkov 61103, Ukraine*

The magnetization anisotropy of a layered superconductor $FeTe_{0.65}Se_{0.35}$ sample is experimentally studied in a magnetic field directed either along the layers of the plane, or perpendicular to them. The value of the vortex pinning potential in a weak magnetic field, and the critical current density ratio are determined for these directions. The results are discussed within the framework of presenting the sample as layers of fine single crystals, divided by weak interlayer superconducting bonds with magnetic inclusions.

## 1. Introduction

The single crystals of iron-based $FeTe_{0.65}Se_{0.35}$ superconductors belong to the family of chalcogenides and have a layered structure. Various authors have studied the superconducting traits of Fe-Te-Se single crystals with this, or a similar composition of elements.[1–5] Thus, as a rule, their transport properties were studied by passing a current through the plane of the *a-b* layers, and the magnetic properties were studied when the magnetic field was applied perpendicular to this plane, i.e., along the *c* axis of the single crystals. Most of the published results come from studies pertaining to the magnetic characteristics of this group of superconductors under strong constant fields (up to tens of T). The possibility of achieving a high transport critical current density ($\approx 10^5$ A/cm$^2$) and the competition of FeTeSe tapes with widely used $Nb_3Sn$ agents during technical use in fields above 20 T,[5] exists.

The goal of this study is to determine how the direction of the magnetic field relative to the *a-b* plane of the single crystal affects its magnetic properties both in the normal and superconducting states, as well as to study the isothermal relaxation of the trapped magnetic field in the given single crystal, which allows us to determine the anisotropy of the vortex pinning potential in a weak magnetic field.

## 2. Experiment Set-Up

The test sample $FeTe_{0.65}Se_{0.35}$ was prepared according to the Bridgman method. The technology used to grow the crystals is described in Ref. 1. The size of the samples was $1.7 \times 1.6$ mm (*a* and *b* axes), with a thickness (along the *c* axis) of $\sim$0.3 mm. The mass of the sample was 5.19 mg. Sample x-ray analysis showed that it had a perfect tetragonal crystallographic structure. The geometric structure of the sample presents a single-layer sandwich made of flat crystallites with an average size of 30 $\mu$m in the *a-b* plane, each of them being a monomolecular layer of the given compound. The crystallites are connected to each other along the *c* axis both mechanically and electrically. The mechanical strength of the connection between the layers is not great, which is confirmed by the fact that we can separate the layers using the Scotch method.[5]

The magnetic properties of the single crystals were examined using MPMS-5 Quantum Design equipment. We obtained sample magnetization temperature dependences in the superconducting transition region, under a weak magnetic field ($H = 5$ Oe), and in the 5–300 K temperature range under strong magnetic fields (up to 500 Oe). In addition, we registered the relaxation of magnetic fluxes in the sample, trapped in a weak (around 5 Oe) constant field, and also the samples' magnetization reversal loops ($H$ of up to $\pm 0.5$ T) at various temperatures. All indicated dependences were obtained for two sample positions relative to the constant applied magnetic field H: for field direction along the *a-b* plane ($H \| a,b$) and perpendicular to this plane ($H \| c$). Before conducting the magnetic measurements on the microbridge, which was cut from a similar single crystal of the same process series, the resistive four-probe method was used to capture the temperature dependences of sample resistivity in the region of the superconducting phase transition (Fig. 1). Given mutually orthogonal directions of the transport current, the critical current density estimates across and along the crystallite layers showed that $J_c(\|c)/J_c(\|a, b) \approx 10$.

Fig. 2 shows the temperature dependence of the sample magnetization $M(T)$ in the region of the superconducting phase transition at different orientations of the sample relative to the direction of the constant magnetic field, equal to 5 Oe.

The magnetic moment can be generated using a non-contact method by exciting the diamagnetic current via the external magnetic field, in the superconducting sample (ZFC-zero field-cooling). We can see from the given curves that the start of the superconducting transition ($T^{onset}{}_c$) gives a value of about 12 K. Concerning the change in the absolute value of magnetization dM during a temperature increase from 5 to 15 K in a 5 Oe field, it differs significantly for two of the orientations of the sample: $dM(H\|c)/dM(H\|a,\ b) \approx 3$. This can be explained by the specific features of the types of possible superconducting circuits with diamagnetic currents that exist in the sample. In the case of $H\|c$, a part of these circuits exists in the crystallites of the sample, located parallel to each other in the sample layers. The perfect microstructure of the sample allows us to assume that the crystallites have a high critical current. The other part of the circuits is composed of the same crystallites and weak interlayer bonds with ferromagnetic inclusions like S-NS,[6,7] therefore having a lower critical current. It is assumed that the planes of these circuits are perpendicular to the *a-b* plane and are therefore parallel to the direction of the magnetic field. Therefore, diamagnetic currents cannot be excited therein, and the corresponding magnetic moment cannot occur.

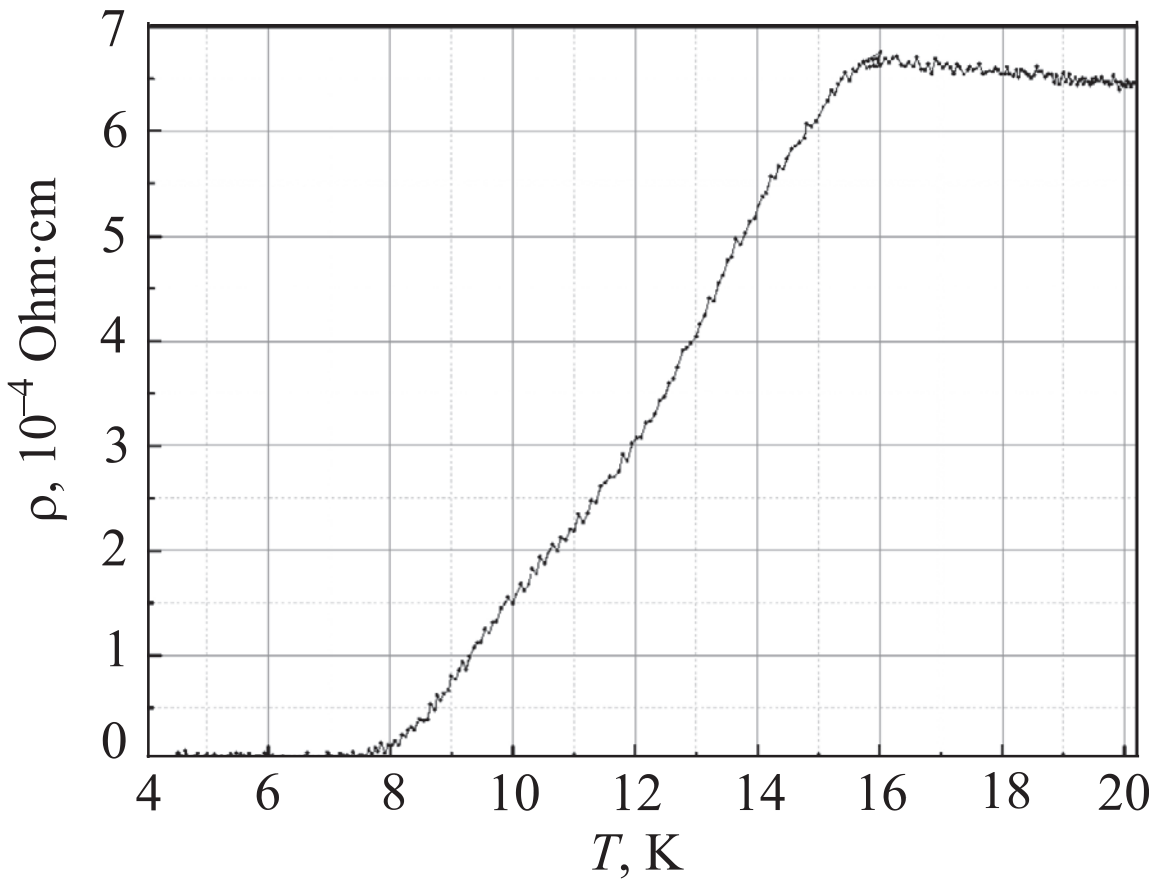

FIG. 1. Temperature dependence of the resistivity ($\rho$) of the sample. The transport current is directed along the *a,b* planes.

In the case of $H\|a,b$ the field parallel to the planes of the monomolecular crystallites, and the size of the diamagnetic circuits in the crystallites, are negligibly small in comparison to the case of $H\|c$. Therefore, their input into the diamagnetic moment is small. Contrary to this, in the other part of the circuits (with weak bonds, and significantly larger in size) at $H\|a,b$ there is a diamagnetic current that creates the magnetic moment of the sample. These circuits occur randomly during the growth of the samples, and their size, number, and the critical current of their weak bonds determine the magnitude of the occurring magnetic moment. As a result we can expect the experiment to give us the relationship between the observed values of the moments at different directions of the field.

Fig. 3 shows the dependence of sample magnetization, normalized to the initial value and caused by trapped vortices in the magnetic field ($H = 5$ Oe) in an FC mode (field-cooling), on time, for one of the experimental temperatures. As can be seen from the presented curves, the field relaxation rate in the case of $H\|a,\ b$ is significantly less than for $H\|c$.

In order to explain the observed difference we will first describe the state of the frozen (trapped) magnetic field in a layered structure of a type-II superconductor, which is consistent with our test material. For the sake of argument, we will once again select a sample model consisting of a set of crystallites with mutually parallel planes, interconnected by weak interlayer superconducting bonds, extended in the direction perpendicular to *a-b* plane of the crystallites. When trapping the field in the FC mode, for $H\|c$, Abrikosov vortices (AV) occur in the crystallites. Thus the circuits with the weak bonds may not contain the trapped field due to the fact that their planes are parallel to the direction of the field. The vortices in the single crystals can be mobilized (in the form of a flux or a jump) by the Lorentz force as well as thermal fluctuations. This process is known as the magnetic flux creep, captured in the crystallites. In particular, the speed at which AV move, which is the relaxation rate of the trapped flux, if any, is determined by the pinning force. In turn, the pinning force is proportional to the pinning potential $U$. The higher the observable relaxation rate, the less the average pinning potential $U$. In particular, for a linear Anderson-Kim relaxation rate (S) model we have[8]

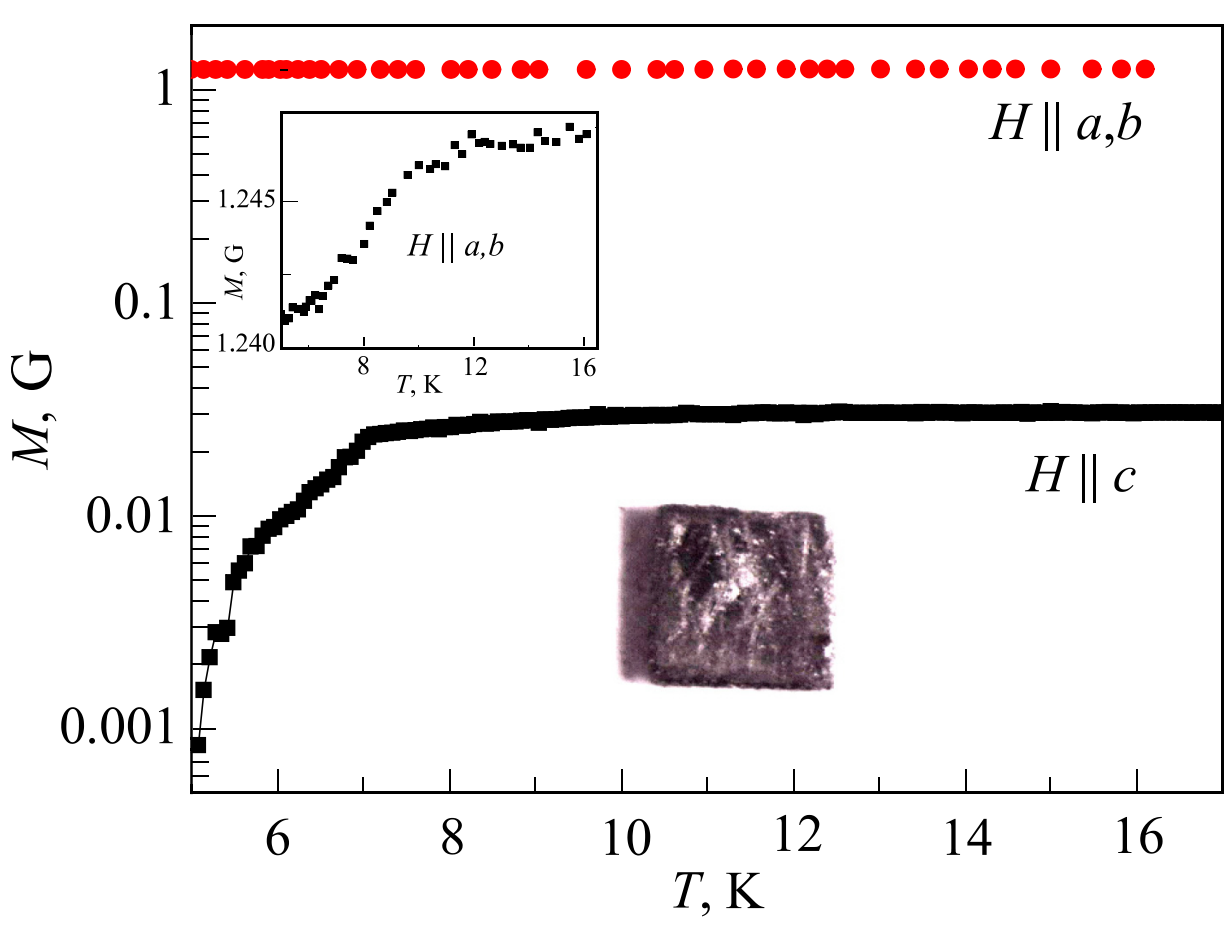

FIG. 2. The temperature dependence of magnetization for the test sample, in the region of the superconducting phase transition at two different orientations of the constant magnetic field ($H = 5$ Oe), and a photo of the crystal (for illustration). The enlarged inset shows a close-up of the initial portion of the $M(T)$ curve, when the field is parallel to the layered structure of the single crystal.

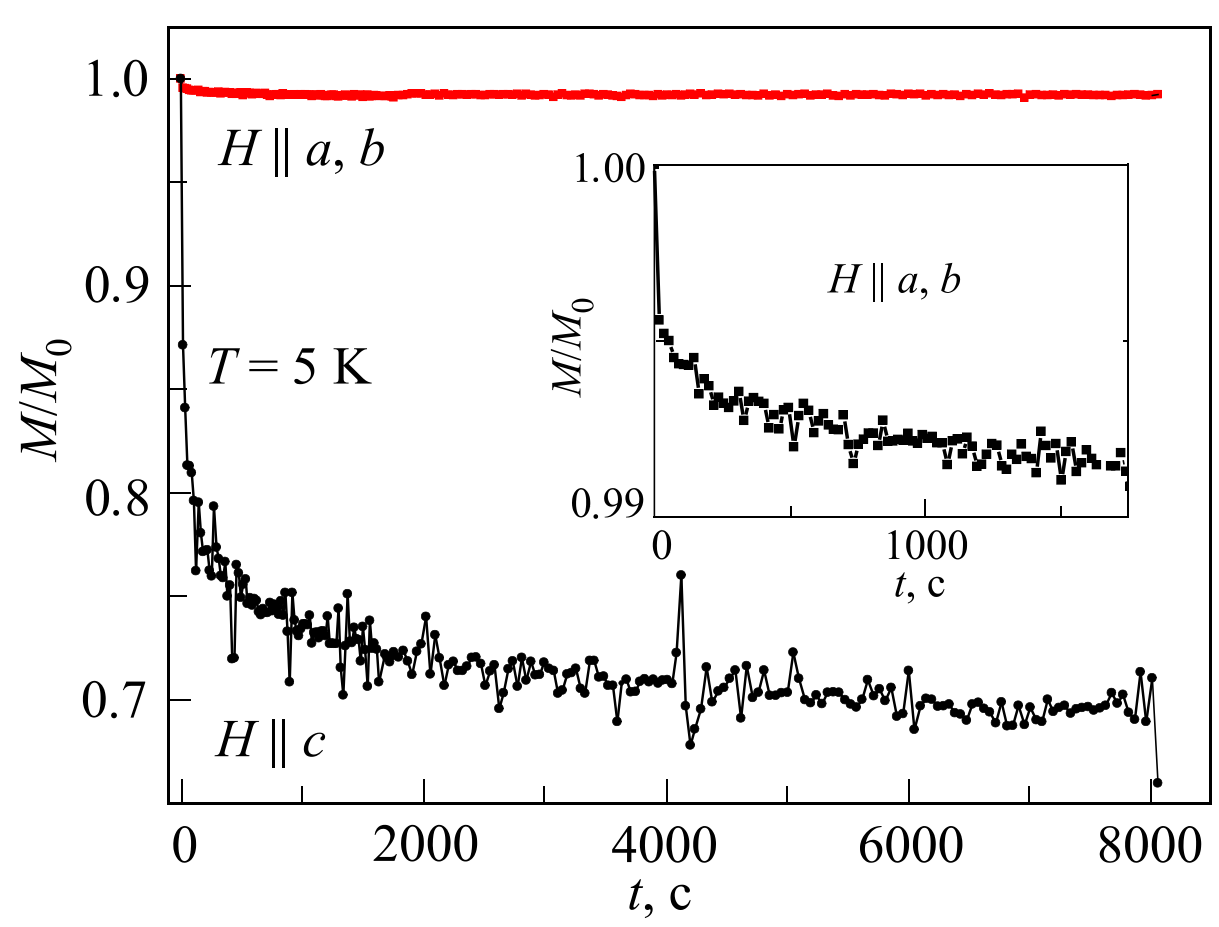

FIG. 3. An example of a typical magnetization relaxation, caused by trapped magnetic flux, pertaining to our test sample at one temperature ($T = 5$ K). The curves $M(T)$ are normalized to the initial value $M_0 = M(t = 0)$ for two of its orientations. The magnified inset shows the initial area of the curve $M(t)/M_0$ at $H\|a,b$.

$$S = \frac{1}{M_0}\frac{dM}{d\ln t} = -\frac{kT}{U}, \qquad (1)$$

where $M_0$ is the initial value of the sample magnetic moment, $k$ is the Boltzmann constant, $T$ is the temperature, and $t$ is time.

In the case of $H \| a,b$ the demagnetization factor of the crystallites is closer to zero because of their small thickness, and the crystallite contribution to the trapped field is small. In contrast to this, the trapped flux input from the magnetic state of the pinning centers and circuits with weak bonds, are the determining factors. In this case we get Abrikosov vortices in regions with a suppressed order parameter, and hyper vortices and Josephson vortices in the weakest links. Assuming that the motion of this type of vortices can also be approximately described by formula (1), we calculated values of $U$ for both direction of the trapped magnetic field (Fig. 4). As can be seen in the figure, the pinning potential of the given compound in the case of $H \| a, b$ is more than an order of magnitude greater than the potential calculated for $H \| c$. As such, due to the layered structure of this type of superconductor, effectively distributed planar pinning centers form for the vortices. It is possible that this is connected to the presence of iron atoms in the interlayer space of the sample.

Fig. 5 shows the magnetization reversal curves of the superconducting sample at $T = 5$ K, in a range of variation for the external magnetic field ±500 Oe, and two directions relative to the *a*-*b* plane.

We can see the manifestation of a strong anisotropy in the crystal's magnetic properties, and a slight deviation from the classical magnetization reversal curves of type-II superconductors.[9] This deviation can be associated with the presence, and the spatial distribution, of magnetic inclusions in the superconducting sample, which exist in this compound and others like it.[10] These inclusions can be $Fe_3O_4$ and $Fe_7Se_8$,[6] located in the interlayer space of the sample. As a result, the total magnetization of the sample $M$ can be represented as two terms

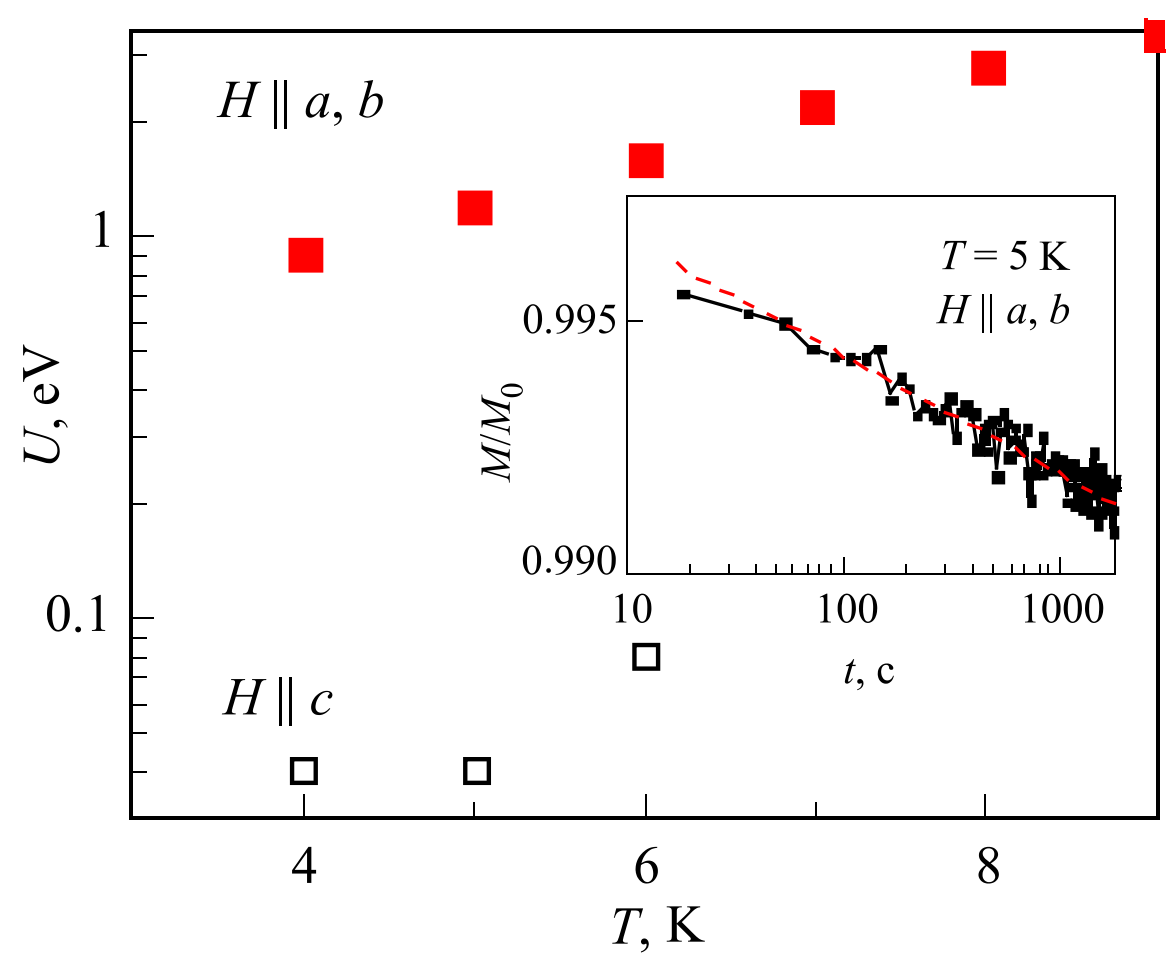

FIG. 4. The average effective pinning potential $U$, according to the volume of the single crystal. This is the range of experimental temperatures for two orientations of the sample ($H \| c$ and $H \| a,b$). The inset shows the experimental data $M(t)/M_0$ for one of the temperatures, on a logarithmic scale. There is a good linear relationship, which allows us to use the Anderson-Kim model to calculate $U$.

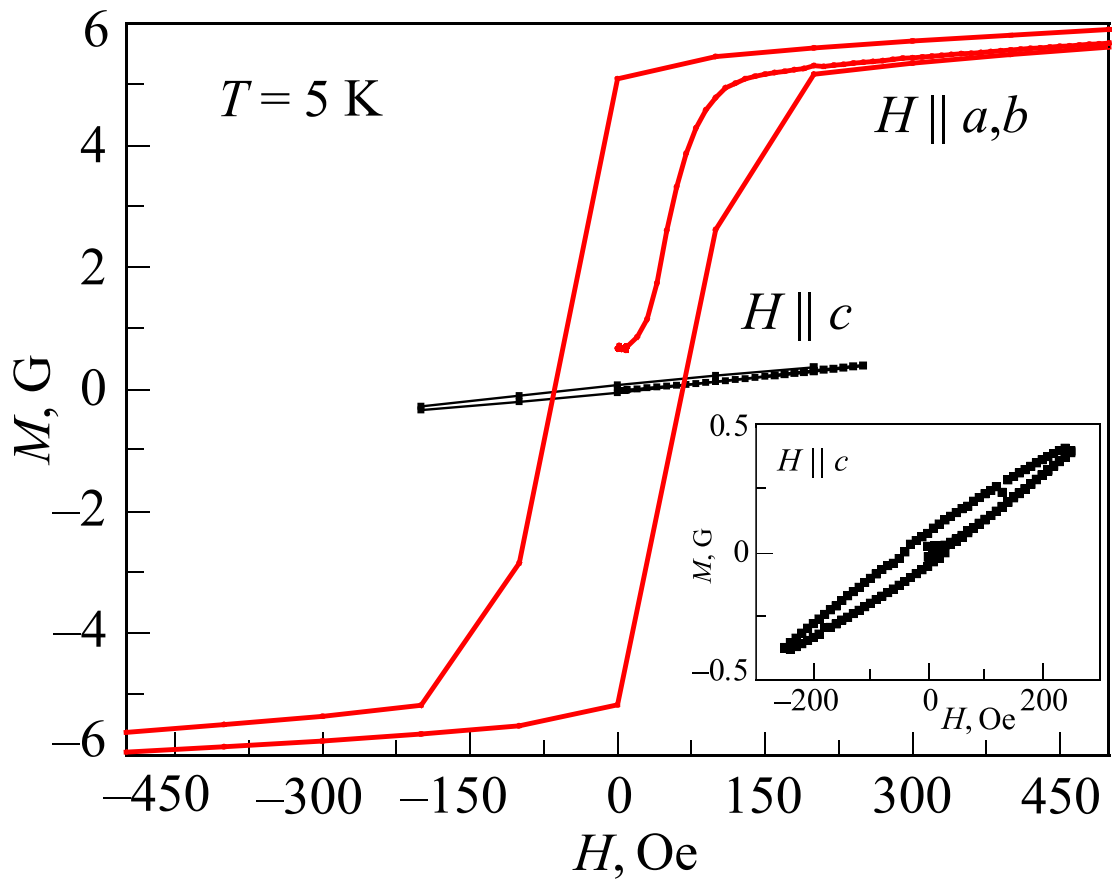

FIG. 5. Magnetization loops $M(H)$ for two sample orientations, at one temperature during the course of the measurements. The magnified inset shows the experimental curve for a case when $H \| c$.

$$M \approx M_f - M_d, \qquad (2)$$

where $M_f$ and $M_d$ are the ferromagnetic and diamagnetic components, respectively. The magnitude of the components depends not only on the value of the magnetic field $H$, but also on the demagnetization factor (shape factor) that depends on the position of the sample relative to the direction of the magnetic field. Analysis of the total magnetization components gives us additional information about the superconducting and magnetic properties of the sample. Let's start with $M_d$. In the range of changes to $H$ that is of interest to us (small fields, temporal order) the input of the ferromagnetic component to the total magnetization can be extrapolated by a linear dependence $M(H)$. After processing the experimental curves using this technique we can estimate the width of the superconductor's magnetization loop that is mainly governed by diamagnetic contribution.

The width of the superconductor's magnetization reversal loop $\Delta M$, that corresponds to the forward and reverse course of $H$ when $M(H)$ is removed, is proportional to the effective depth of the pinning potential $U$, averaged over the sample volume. According to Bean's critical state model, the critical current density of the superconductor $J_c$ is associated with the geometric parameters of the test samples, and $\Delta M$. The critical current density $J_c$ can be estimated according to the famous formula $J_c = 20\Delta M/[a(1 - a/3b)]$, wherein $a, b$ $(a < b)$ are the cross-section dimensions of the sample.[11] The relationship of the experimental values $\Delta M(H \approx 0)$ obtained via this method, in the mutually perpendicular orientations of the single crystals, $\Delta M \approx 1.2$ G ($H \| a, b$) and $\Delta M \approx 0.13$ G ($H \| c$), gives us the quotient

$$\frac{J_c(H \parallel a, b)}{J_c(H \parallel c)} \approx 20, \qquad (3)$$

which is in qualitative agreement with the data shown in Fig. 4, for the effective pinning potential.

We will now examine the $M_f$ component. If we assume that the excess iron impurity and its oxides form something resembling a magnetic layer between the crystallites, then the observed magnetization reversal curves can be explained by the different coefficients of demagnetization for these layers, under a magnetic field along and across the *a*-*b* plane of the sample. Along the layers this coefficient is close to zero, whereas across, it is equal to one. As a result, the direction across the *a-b* plane is the sample's easy axis of magnetization, with a correspondingly narrow magnetization reversal curve, and is transverse to the plane, unlike the field direction.

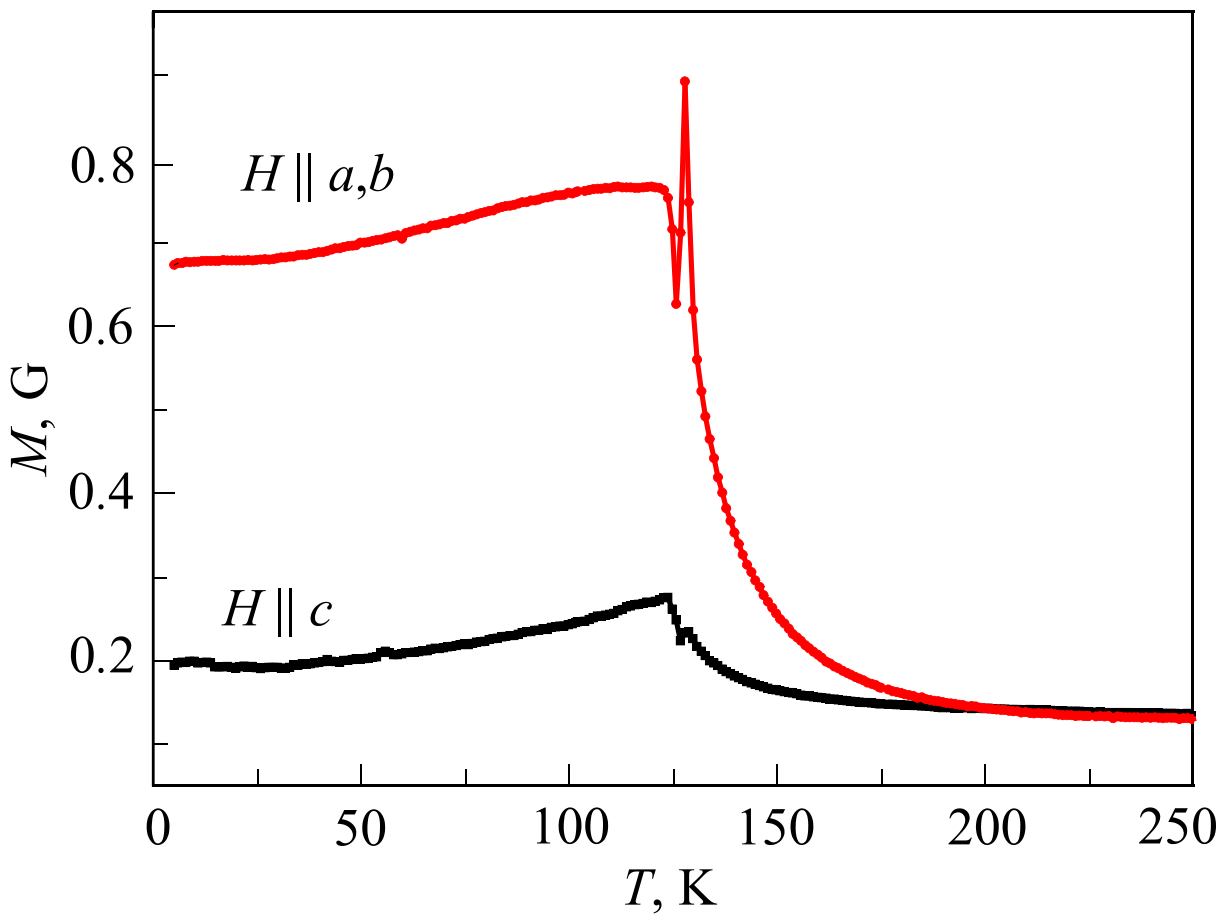

FIG. 6. The magnetization of the experimental single crystal sample for two of its orientations under the magnetic field $H = 5$ Oe, in an extended temperature range, which demonstrates the influence of the magnetostructural transition of the crystal lattice.

Fig. 6 shows the temperature dependence of the sample's magnetic moment in an expanded temperature range (including the normal state of the crystal) at two directions of the magnetic field.

The expected difference in the amplitude of magnetization and the agreement in the position of the jump $M$ at $T \approx 125\,K$, are both registered. The jump is observed earlier by other researchers in much stronger fields and is associated with the presence of a magnetostructural transition in the layers of the magnetic impurity $Fe_3O_4$.[6] The difference in amplitude can be explained similar to the difference in the $M_f$ component in the total magnetization, as discussed above.

## 4. Conclusions

Magnetic moment measurements of superconducting chalcogenide $FeTe_{0.65}Se_{0.35}$ under a magnetic field directed along its layers and across, serve as evidence of a significant dependence of the material's magnetic properties both in the superconducting and in the normal state, on the direction of the field. It is found that a decisive role in the manifestation of this dependence is played by the features of sample's crystal structure. The test crystal is composed of fine (around 30 $\mu$m in *a-b* layer planes) flat single $FeTe_{0.65}Se_{0.35}$ crystallites that are parallel to each other, and divided by interlayer superconducting weak S-N-S type bonds. A feature of these bonds is the complex composition of the normal (N) layer, presumably containing the magnetic phases $Fe_3O_4$ and $Fe_7Se_8$.

The measured values of the diamagnetic and paramagnetic moments and their dependence on the direction for the field, are determined by the superconducting and magnetic properties of two types of superconductive current circuits: current loops appearing in the single crystallites themselves (in the form of Abrikosov vortices), and current loops formed by the single crystallites and the weak bonds between them. At the same time, in our opinion, the planes of these two types of circuits are mutually perpendicular. Therefore for one direction of the field that determines the value of the magnetic moment, there is a set of circuits of the same type, whereas for another direction the magnetic moment is determined by the properties of the aggregate of another type of circuit.

This feature of the circuits allowed us to conduct a diagnostic analysis of the properties of the two types of sample environments (the layers of the $FeTe_{0.65}Se_{0.35}$ single crystals, and the magnetic interlayers) by measuring the magnetization of the sample in the two orthogonal external fields of current excitation in the sample. In particular, as a result of the magnetization relaxation measurements that were conducted, we were able to calculate and demonstrate the magnitude and temperature dependence of the vortex pinning potential in a weak magnetic field (5 Oe), and establish the value of the anisotropy for two field directions, as well as calculate the critical current density ratio at these directions.